\documentclass[traditabstract]{aa}

\usepackage{graphicx}   
\usepackage{txfonts}
\usepackage{amstext}
\usepackage[dvipsnames]{xcolor}
\usepackage{hyperref}
\hypersetup{
 colorlinks   = true, 
 urlcolor     = blue, 
 linkcolor    = blue, 
 citecolor    = blue  
 }

\newcommand{\ks}{km~s$^{-1}$}
\newcommand{\ms}{M$_{\odot}$}
\newcommand{\rs}{R$_{\odot}$}
\newcommand{\oc}{$O\!-\!C$ }
\newcommand{\hip}{{\sc Hipparcos}}
\newcommand{\T}{TESS}

\begin{document}

\title{Apsidal motion and \T\ light curves of two southern eclipsing binaries with high eccentricity: V1647~Sgr and V2283~Sgr
\thanks{Based on observations secured at the 
        South Africa Astronomical Observatory, Sutherland, South Africa,
        La Silla observatory, Chile,
        FRAM, Pierre Auger Observatory, Argentina, and
        Boyden observatory, South Africa.
    Tables~A1 -- A3 are available in the CDS via anonymous ftp to \url{cdsarc.cds.unistra.fr} (130.79.128.5) or via \url{https://cdsarc.cds.unistra.fr/viz-bin/cat/J/A+A/???/A??} \\
        }  }


\author{M. Wolf~\inst{1}
         \and P. Zasche~\inst{1}
         \and M. Zejda \inst{2} 
         \and M. Ma\v{s}ek \inst{3, 4}
        }

   \institute{Astronomical Institute, Faculty of Mathematics and Physics,
   Charles University Prague, V Hole\v{s}ovi\v{c}k\'ach 2, \\
   CZ-180~00 Praha 8, Czech Republic, E-mail: {\tt marek.wolf@mff.cuni.cz}   
   \and Department of Theoretical Physics and Astrophysics, Masaryk
   University, Kotl\'a\v{r}sk\'a 2, CZ-611~37 Brno, Czech Republic 
   \and FZU - Institute of Physics of the Czech Academy of Sciences, Na Slovance 1999/2, 
   CZ-182~21, Praha 8, Czech Republic
   \and Czech Astronomical Society, Variable Star and Exoplanet Section, V\'ide\v{n}sk\'a~1056, CZ-142~00~Praha~4, Czech Republic
   }

\date{Accepted XXX; Received YYY; in original form ZZZ}




\abstract{
The study of apsidal motion rates in eccentric eclipsing binaries provides
an important observational test of theoretical models of stellar structure and evolution. Precise physical parameters of the stellar components together with systematic measurements of the periastron advance are needed.
We present new results of our long-term observational project to analyze the apsidal motion in early-type eccentric eclipsing binaries.
New ground and space-based photometric data were obtained, and archival spectroscopic measurements were used in this study of two detached southern-hemisphere eclipsing binaries:  
   V1647~Sgr  ($P = 3\fd28, e = 0.41$), and
   V2283~Sgr  ($3\fd47, 0.49$).
 Their \T\ observations in four sectors have also been included and the corresponding light curves were solved using the {\sc Phoebe} code. 
 The newly completed \oc\ diagrams were analyzed using all reliable timings found in the literature and calculated using the \T\ light curves. 
 New or improved values were obtained for the elements of apsidal motion. 
 Using archival spectroscopy
 for V1647~Sgr, the precise absolute parameters were improved:  
 $M_1$ =   2.184(0.035)~\ms,  
 $M_2$ =   1.957(0.035)~\ms, and
 $R_1$ =   1.839(0.015)~\rs,  
 $R_2$ =   1.716(0.015)~\rs.
 For V2283~Sgr the absolute dimensions were newly estimated:
 $M_1$ =   2.178(0.10)~\ms, 
 $M_2$ =   1.547(0.10)~\ms, and
 $R_1$ =   1.796(0.01)~\rs,  
 $R_2$ =   1.544(0.01)~\rs.
We improved relatively long periods of apsidal motion of about
580 and 530 years, together with the corresponding internal structure constants, 
log $k_2$, --2.394, and --2.418, for V1647~Sgr and V2283~Sgr, respectively.
The relativistic contribution to apsidal motion is not negligible, making about 12 resp. 14\% of the total rate of apsidal motion. 
No signs of the presence of an additional body were revealed in the light curves or in the \oc\ diagrams of both eccentric systems. }

\keywords{
binaries: eclipsing --
binaries: close --
stars: early-type --
stars: fundamental parameters --
stars: individual: V1647~Sgr, V2283~Sgr}

\authorrunning{M. Wolf et al.}
\titlerunning{Apsidal motion and \T\ light curves of V1647~Sgr and V2283~Sgr}

\maketitle

\section{Introduction}

\begin{table*}
\caption {Properties of both selected eccentric eclipsing binaries.
 The spectral type, $V$ and $G$ magnitudes, and parallax values  
 are taken from the {\sc Simbad} database. }
\label{summ}
\begin{tabular}{cllccccccl}
\hline\hline\noalign{\smallskip}
System & HD    & CD   & Spectral & Period & $V$  & $B-V$ & $G$  &  Parallax & Eccentricity\\
       & number& number    & type      &  [day] & [mag]& [mag] & [mag]&   [mas]  &  -- \\
\hline\noalign{\smallskip}
V1647~Sgr& HD~163708 & CD-36~12064 & A1V & 3.283 & 7.09 & 0.066 & 7.071 & 6.01 & 0.412 \\
V2283~Sgr& HD~321230 & CD-36~12180 & A0V & 3.471 & 10.4 & 0.06 & 10.282 & 1.62 & 0.489 \\
\hline\noalign{\smallskip}
\end{tabular}
\end{table*}

Some bright stellar objects still escape modern study, but they are of fundamental importance for a broader astrophysical context in stellar astrophysics. One such group is eccentric eclipsing binaries (EEB) which provide an ideal opportunity to verify stellar structure as well as to test General Relativity effects outside of the Solar System \citep{2007IAUS..240..290G, 2010A&A...519A..57C, 2019ApJ...876..134C}.
Moreover, detached double-lined eclipsing binaries (DLEBs) simultaneously serve as an important source of fundamental information on stellar masses and radii
\citep{1991A&ARv...3...91A}. 
\cite{2010AARv..18...67T}  revised the absolute dimensions of 94~DLEBs and only 18 of them are found to have eccentric orbits that show apsidal motion suitable for the apsidal motion test of stellar structure.

Some time ago, an analysis of apsidal motion in eclipsing binaries using \T\ data was presented by \cite{2021A&A...649A..64B, 2022A&A...665A..13B}. 
These authors determined the apsidal motion rate for nine EEBs, measured the general relativistic apsidal motion rate, and performed a test of general relativity.
They found perfect agreement with theoretical predictions and established stringent constraints on the parameters of the post-Newtonian formalism.
Recently, \cite{2021A&A...654A..17C} studied the internal structure constant
in 34~EEBs and nearly doubled the sample of suitable systems collected by \cite{2010AARv..18...67T}. Their comparison of the apsidal motion rate with predictions based on new theoretical models shows very good agreement. 
Therefore, it is necessary to expand the current collection of EEBs with precise absolute parameters to provide better statistics on these results.
The only exception is the EM~Car with the highest mass of both components
(22.8 + 21.4 \ms), which deserves further attention \citep{2021A&A...654A..17C}. 

However, a precise determination of the apsidal motion rate requires 
long-term monitoring of mid-eclipse times, usually spanning several decades, 
and many advanced amateur observers can help with this difficult task. 
Apsidal motion studies and the determination of precise stellar parameters 
are still often the subject of many papers.
Let us recall the extensive studies of \cite{2016MNRAS.460..650H} and
\cite{2020A&A...640A..33Z} or recent works on individual objects
\citep{2024Obs...144..278S, 2024PZ.....44..131K, 2024A&A...683A.158Z, 2025ARep...69..480V}.
The apsidal motion rate can also be determined with relatively good precision in spectroscopic binaries of the SB2 type, where the radial velocities of the primary and secondary components are measured over a long time interval 
 \citep{2016A&A...586A.104S, 2016A&A...594A..33R} 
by changes in the shape of the light curve over time 
\citep{2014AA...563A.120H, 2017ApJ...836..177T},
 or even by combining photometry and spectroscopy together
\cite{2020A&A...635A.145R, 2022A&A...660A.120R, 2022A&A...664A..98R}.

Both binary systems under study have many similarities and common characteristics:
a rather longer orbital period of about 80 hours, early spectral type A, high orbital eccentricity (more than 0.4), slow apsidal motion with a similar period of over 500 years, and finally, their close position in the southern sky in the Sagittarius constellation ($\sim$ two diameters of the Moon apart).  
Recall that probably the highest known eccentricity $e=0.618$ was discovered in the eclipsing binary V680~Mon by \cite{2021ARep...65..184V}. 
Their basic properties are summarized in Table~\ref{summ}.
This study of apsidal motion in EEBs is a continuation
of our work presented in earlier papers over several decades, 
the last time in \cite{2022NewA...9201708W,2024AA...690A.231W}.
The current paper is organized as follows. 
Section~\ref{observ} deals with new observations and data reductions. 
The apsidal motion analysis is given in Sect.~\ref{apsmotion}, 
while the light curves are analyzed in Sect.~\ref{lcrv}. 
The internal structure constants are calculated in Sect.~\ref{ISC} 
and a brief summary of the results is presented in Sect.~\ref{Concl}.

\medskip
The double-lined eclipsing binary V1647~Sgr (also HD~163708,
$V_{\rm max}$ = 7.09~mag) is a bright A-type binary
with a high orbital eccentricity ($e$ = 0.41). 
It is located in the vicinity of the bright G8 giant star HD~163652 ($V = 5.7$~mag).
Since its discovery by J. Herschel, V1647~Sgr is known as a primary component
of the visual double star HJ~5000A.
It was discovered to be a variable star by \cite{1956BAN....12..338P} and independently studied by \cite{1955RA......3..277D}. 
The complete simultaneous Stromgren four-color photometry of V1647~Sgr 
was obtained in 1973-74 and 1982 at ESO in Chile \citep{1977A&A....58..121C}. 
The spectroscopic orbital elements and absolute dimensions were later determined
by \cite{1985AA...145..206A}. They also derived an apsidal motion with a period 
of $U = 592.5\pm 6.5$ yr and a substantial value for the orbital eccentricity
of $e$ = 0.4130. In that paper, the following linear ephemeris was derived:

\begin{center}
Pri. Min. = HJD 24 41829.69510(1) + $ 3\fd28279251(1) \cdot E,$ \\
Sec. Min. = HJD 24 41830.55561(8) + $ 3\fd28282227(33) \cdot E.$
\end{center}

\noindent
\cite{2000AA...356..134W} in his study of apsidal motion confirmed the parameters
derived earlier with a rather shorter apsidal motion period $U = 531 \pm 5$ years.
V1647~Sgr was included in the recent study of \cite{2021A&A...654A..17C}, where the observed and theoretical internal structure constants of 27 EEBs were compared.  
The comparison of observed ISC with the predictions based on new theoretical models 
shows very good agreement.
For V1647~Sgr they found $\log k_{\rm 2,obs} = -2.373$ and $\log k_{\rm 2,theo} = -2.384$ in good agreement between theory and observation. 

V1647~Sgr was also incorporated into the study of \cite{2021A&A...647A..12K} among 
the so-called 'heartbeat stars', which are eccentric binaries that exhibit
a characteristic shape of brightness changes close to the periastron passage,
primarily caused by variable tidal distortion of the components.
\cite{2023AJ....166..114D} inserted this object among 15 eccentric eclipsing binaries known for their apsidal motion. They compared the derived apsidal rates with an analytic solution, resulting in good agreement.

\medskip
\noindent

In contrast, the detached eclipsing binary V2283~Sgr (also HD 321230,
$V_{max}$ = 10.4~mag; Sp. A0) is a photometrically and spectroscopically neglected binary with a remarkably eccentric orbit ($e$ = 0.49) and a period of about 3.5 days. It was discovered to be a variable by Ms. Henrietta Swope on photographic 
plates in MWF~187 \citep{1938BHarO.909....5S}.      
\cite{1965AnLei..22...91K} rediscovered this star
in the region around Boss~4599 and found a very high orbital eccentricity 
of at least 0.45. 
The first photoelectric observations she obtained by \cite{1974RA......8..481S} at the Siding Spring observatory in Australia in 1965, where she derived the following linear light elements: 

\begin{center}
Pri. Min. = HJD 24 38948.5043 + $ 3\fd4714231 \cdot E,$ \\
Sec. Min. = HJD 24 38946.7619 + $ 3\fd4714231 \cdot E,$
\end{center}

\noindent
and the substantial value of $e$ = 0.49 for the orbital eccentricity.
Swope also found an apsidal motion with a period of $U$ = 570 years.
Independently, \cite{1974RA......8..491O} derived similar results, 
eccentricity 0.487 and an apsidal motion period of 560~years.
Finally,  \cite{2000AA...356..134W} confirmed the parameters mentioned above
with a rather shorter apsidal motion period of $U = 528 \pm 12$ years.
This binary has rarely been investigated since its discovery, and no spectroscopic observations have been published for this system as far as we know.

Our last study on these two southern EEBs \citep{2000AA...356..134W} was presented 
25~years ago, so we decided to recalculate the apsidal motion with newly available data.
In particular, very precise measurements from the \T\ satellite are available,  
and we have a number of new mid-eclipse times obtained at several southern 
observatories. 

\section{Observations}
\label{observ}
\subsection{Ground-based photometry}

Our new photoelectric and CCD observations were obtained at four 
southern observatories in the past, in chronological order: 

\begin{description}

\item{$\bullet$} South African Astronomical Observatory (hereafter SAAO), 
Sutherland, South Africa: the 0.50-m Cassegrain reflector ($f/18$) equipped
with a modular photometer utilizing a Hamamatsu EA1516 photomultiplier and
Johnson \textit{UBV} filters in September 2005. 
For bright V1647~Sgr a neutral density filter ND1 was included. Alternatively, a Helios 2/58 objective with an ATIK 16 IC CCD camera; during two weeks in November 2008 and April 2010. The Helios 2/58 objective with a G2-402 CCD camera (\textit{BVR} filters) was also used in March~2018. 

\item{$\bullet$} La Silla Observatory in Chile: Danish 1.54-m reflecting telescope, EFOSC camera, and Johnson \textit{B} filter, remote control during one night on February 17,~2016. 

\item{$\bullet$} Fotometric Robotic Atmospheric Monitor \citep[FRAM]{2021JInst..16P6027A}, Pierre Auger Observatory, Argentina: 
telephoto lens Nikkor 300~mm ($f/2.8$), CCD camera G4-16000 and attenuating filter or 
ODK 0.30-m telescope ($f/6.8$), CCD camera G4-16000 and {\it R} filter.  
Remote control during two epochs in May/August 2020 and July~2025.

\item{$\bullet$} Boyden Observatory, South Africa: reflecting telescope Celestron CGE 1400 XLT (350/2250) and C3-26000Pro CMOS camera, and Sloan  \textit{gri} filters;  in remote control during several nights in June~2025.

\end{description}

\noindent
The main aim of these photometric measurements was to secure several well-covered primary and secondary eclipses for both variables. Each observation of an eclipsing binary was accompanied by the observation of local comparison and check stars (see Table~\ref{comp}).  

\begin{figure}[t]
\centering
\includegraphics[width=0.47\textwidth]{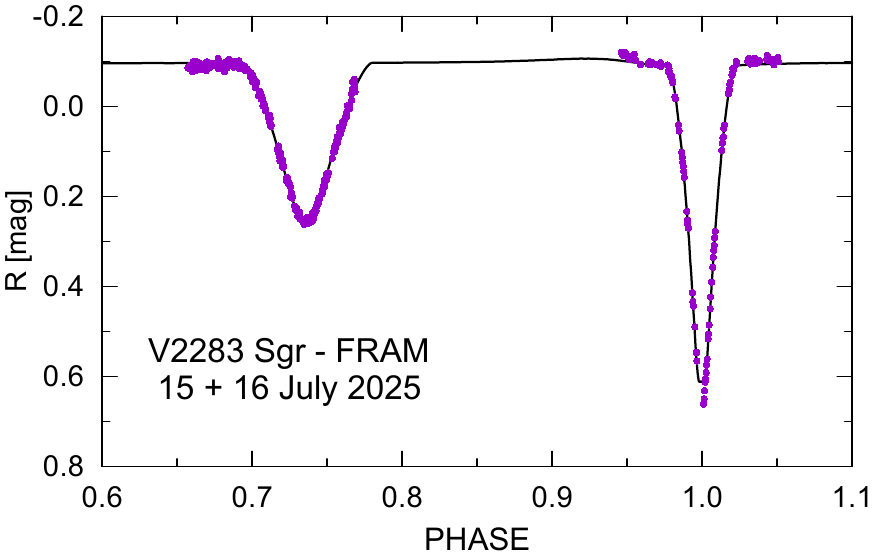}
\caption[ ]{Example of our photometric observations of V2283~Sgr 
obtained at FRAM observatory, Argentina, during two consecutive nights 
in July 2025. Differential photometry in $R$ filter and its fit in {\sc Phoebe}. 
The secondary minimum of V2283~Sgr falls close to phase 0.75.}      
\label{fram}
\end{figure}

Photoelectric measurements at SAAO were usually performed using Johnson's \textit{UBV} photometric filters with a 10-second integration time.  
All observations were carefully reduced to the Cousins E~region standard
system \citep{1989SAAOC..13....1M} and corrected for differential extinction
using the reduction program HEC~22 rel.~16 \citep{1998JAD.....4....5H}.
The standard errors of the \textit{UBV} measurements at SAAO were 0.011, 0.008 and 0.007~mag in the $U, B$ and $V$ filters, respectively. 
The {\sc C-Munipack}\footnote{Motl, 2018, \url{https://c-munipack.sourceforge.net/}, ver. 2.1.24} software package for aperture photometry, 
based on the {\sc Daophot} procedure, was used to process the set of CCD frames. 
An example of our photometric observations of V2283~Sgr obtained at FRAM
observatory in July 2025 is given in Fig.~\ref{fram}.

\subsection{\T\ photometry}
Moreover, the Transiting Exoplanet Survey Satellite (TESS) mission
to study exoplanets through photometric transits \citep{2015JATIS...1a4003R}, 
with its nearly full sky coverage, also provides precise photometry of a large
sample of eclipsing binary systems with a time baseline of 27 days to several years. 
Thus, precise monitoring of light curves is possible from space, and their exceptional
quality is perfect for studying light curves and eclipse timings \citep{2021A&A...649A..64B}. 
Both systems studied in this work have been observed by \T\  in four identical sectors during 2019 - 2025 (see Table~\ref{tess}).  For V1647~Sgr and V2283~Sgr, we collected simple aperture photometry (SAP) of the available cadence produced by the Science Process Operation Centre (SPOC) \citep{2016SPIE.9913E..3EJ} available at 
Mikulski Archive for Space Telescopes (MAST)~\footnote{\url{https://mast.stsci.edu/portal/Mashup/Clients/Mast/Portal.html }}. 

The pixels of the \T\ detectors cover a sky area of 21 × 21 arcsec. 
The \T\ aperture photometry is extracted over several pixels, thus we checked the Gaia catalogue \citep[DR3]{2022yCat.1355....0G} for possible contamination of our objects. 
The catalog lists up to a thousand objects (in Sgr), all of which are significantly fainter than our targets. 
The brightest nearby star of V1647~Sgr is a dominant close visual companion 
HD~163708B $\sim$ 2~mag fainter. 
In the case of V2283~Sgr, there is a 3~mag fainter star, Gaia DR3 4037120519350665472.
Hence, formally, we included the third light in our {\sc Phoebe} solution
to reduce contamination from nearby sources.

\begin{table}
\caption{The common \T\ visibility of V1627~Sgr and V2283~Sgr 
and sectors used for light curve analyses and for mid-eclipse 
time determination.}
\centering
\label{tess}
\begin{tabular}{ccc}
\hline\hline\noalign{\smallskip}
 Sector  &  Start date  & Exposure     \\
   No.   &  YYYY-MM-DD  & time [sec]  \\
\hline\noalign{\smallskip}
    13  &  2019-06-19  & 1800  \\ 
    39  &  2021-05-27  &  600  \\  
    66  &  2023-06-02  &  200  \\    
    93  &  2025-06-03  &  158   \\           
\hline\noalign{\smallskip}
\end{tabular}
\end{table}

The new times of primary and secondary minima and their errors
were determined using a least-squares fit of the light curve during eclipses. 
The third- and fourth-order polynomial fittings were applied to the 
bottom sections of the light curves, and the mean value was adopted as mid-eclipse times. 
Where needed, the mean values of the individual filter bands 
are given, and the presented errors represent the fitting mean error 
for each light curve. 
For the \T\ data, the mid-eclipse times were determined using the
{\sc Silicups}~\footnote{SImple LIght CUrve Processing System, \\
\url{https://www.gxccd.com/cat?id=187&lang=405}} code.
Because the \T\ data are provided in the Barycentric Julian Date Dynamical Time  
(BJD$_{\rm TDB}$), all our previous times were first transformed to this time
scale using the often used Time Utilities of Ohio State 
University~\footnote{\url{http://astroutils.astronomy.ohio-state.edu/time/} } 
\citep{2010PASP..122..935E}.
All new minima times for V1647~Sgr and V2283~Sgr in BJD are collected in Tables~\ref{minground} and \ref{mintess}, and the epochs are calculated from the 
light ephemeris given in the text.

\subsection{ASAS-3, OMC, and Pi of the Sky photometry}

Using the All Sky Automated Survey-3 (ASAS-3) database~\footnote{\url{http://www.astrouw.edu.pl/asas/}} \citep{2002AcA....52..397P}, 
photometric data from the {\sc OMC Integral} Optical
Monitoring Camera Archive 
\citep{2004ESASP.552..729M}~\footnote{\url{https://sdc.cab.inta-csic.es/omc/}},
and Pi of the Sky \citep{2005NewA...10..409B},
we were able to derive several additional times of minimum light with a precision
of about 0.001-0.002 days. They were used in our next analysis with a weight of 1.

To derive some of the eclipse times from various surveys with sparse photometry, 
we used our method named AFP (described in \cite{2014A&A...572A..71Z}). It uses the phased light curve over a longer time interval and the template of the light curve provided through light curve modeling in {\sc Phoebe}.
The time interval used was typically one year, but this can be arbitrarily changed with respect to the number of data points in each interval. We are typically able to derive several times of eclipses, primary and secondary, using this method from one particular photometric survey dataset.

\subsection{Spectroscopy}

The previously published spectroscopy and radial velocity curve are available 
only for V1647~Sgr \citep[their Table 2]{1985AA...145..206A}. 
A total of 27~spectrograms were obtained at the 1.5-m ESO telescope and coudé spectrograph at La~Silla, Chile, in two consecutive seasons in 1976-1977. 
We used this precise and compact dataset for a common solution in {\sc Phoebe}.
To our knowledge, no spectroscopic material or radial velocities have been 
reported in the literature for the relatively bright target V2283~Sgr, nor found
in any available electronic database.


\section{Apsidal motion analysis}
\label{apsmotion}

\begin{figure}[t]
\centering
\includegraphics[width=0.47\textwidth]{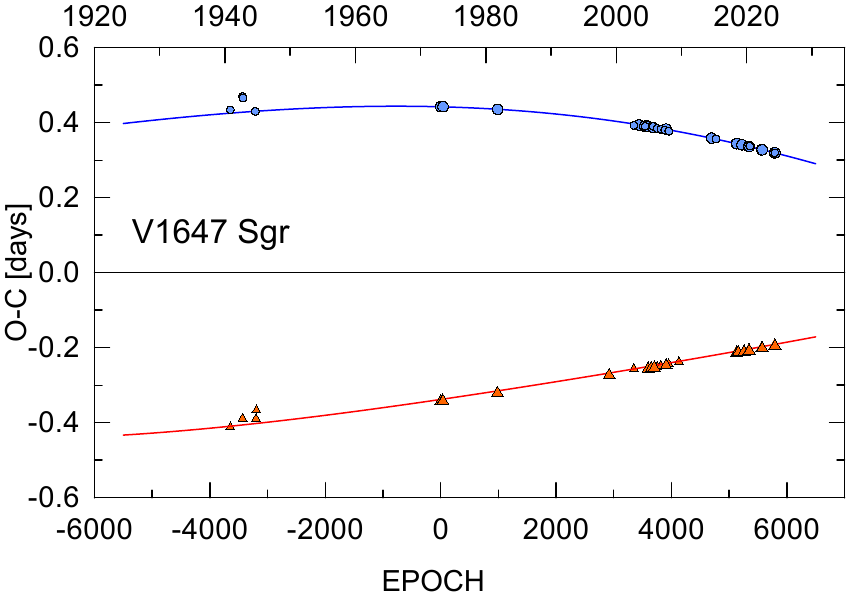}
\caption[ ]{Complete \oc\ diagram of V1647~Sgr spanning almost one century. Primary minima are denoted by blue circles, secondary by orange triangles. The curves correspond to our best-fit apsidal motion model. All TESS data represents four clusters of points at the end of both curves.}      
\label{1647oc}
\end{figure}

\begin{figure}[t]
\centering
\includegraphics[width=0.5\textwidth]{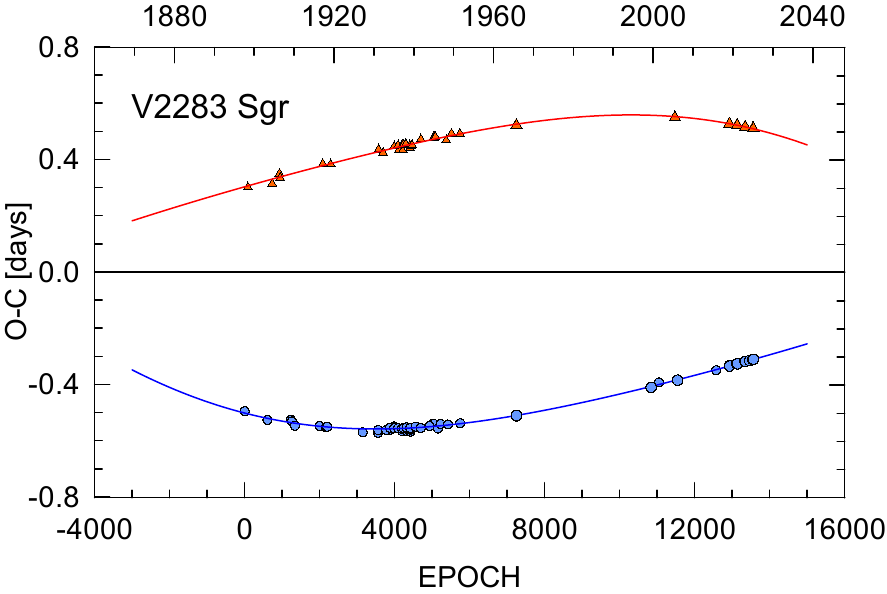}
\caption[ ]{Historical \oc\ diagram of V2283~Sgr since the beginning of 20th century together with our best-fit apsidal motion model. 
See legend to Fig.~\ref{1647oc}.        } 
\label{2283oc}
\end{figure}

The apsidal motion study requires long-term monitoring of eclipse times, usually spanning several decades. This is not an observing program that requires large telescopes, but advanced amateur astronomers can help with this time-consuming task.
As in our previous studies, apsidal motion was studied using an 
\oc\ diagram (or ETV~curve) analysis. 
The iterative method described by \cite{1983Ap&SS..92..203G} or \cite{1995Ap&SS.226...99G} was used again.
We collected all suitable mid-eclipse times available in the literature and current databases. The input files of the minima were completed with our own measurements and all available data from \T.
All photoelectric and CCD times, as well as the \T\ minima were used with a weight of 10 in our calculation. Elderly, less accurate measurements with unknown uncertainty (photographic or visual estimations) were assigned weights 1 or 0.
However, due to the long time lag and large dispersion of these measurements from the first half of the 20th century, the final solution may depend on the adopted weighting scheme.
The apsidal motion parameters were calculated iteratively. For the orbital inclination, we adopted the results of our own photometric solution
(see Chapter~\ref{lcrv} and Table~\ref{lcfit}). On the contrary , the eccentricity resulting from the analysis of the apsidal motion was fixed in the following solution of light curves. In general, the apsidal motion solution was found to be insensitive to relatively large changes in inclination but strongly depended on the orbital eccentricity.

In the case of V1647~Sgr, original photographic times of the minima of \cite{1956BAN....12..338P} and all the photoelectric times given in  
\cite{1977A&A....58..121C} (their Tables~2 and 3) were incorporated into our calculation. Several times with less precision were found in the literature
or derived using Pi of the Sky and OMC photometry. 
Our new $VR$ CCD photometry obtained at the FRAM and Boyden observatories
was also used to determine new mid-eclipse times. 
A large number of very precise and consecutive times were obtained using
the \T\ data in four sectors. 
A total of 94 times of minimum light were used in our analysis, with 48 secondary eclipses among them. All are listed in Tables~\ref{minground} and \ref{mintess}.

For V2283~Sgr there is a limited number of new precise electronic mid-eclipse times.
All photographic eclipse times given in \cite{1938BHarO.909....5S} and
\cite{1974RA......8..491O} were included in our analysis. 
Four new and precise times were derived using our FRAM and Boyden observations.
The most numerous contribution of minima comes from four \T\ sectors. 
The newly determined times of minimum light are listed in Tables~\ref{minground} and \ref{mintess}, where all epochs were computed according to the ephemeris given in the text. In our analysis of V2283~Sgr, a total of 115 times of minimum light were used.

The residuals of \oc\ values for all minimum times with respect to the
linear part of the apsidal motion equation are shown in Fig.~\ref{1647oc}
and Fig.~\ref{2283oc}.
The computed elements of apsidal motion and their internal errors
of least squares fit are given in Table~\ref{t2},
where $\dot{\omega}$ is the rate of periastron advance
(in degrees per cycle or in degrees per year), and the periastron position for
the zero epoch $T_0$ is denoted as $\omega_0$. 
The relation between the sidereal and the anomalistic period, $P_{\rm s}$ and $P_{\rm a}$, is given by

\begin{equation}
    P_{\rm s} = P_{\rm a} \,(1 - \dot{\omega}/360^\circ), 
\end{equation}

\noindent
which means that the sidereal period is always a little shorter than the anomalistic period due to apsidal motion. Finally, the period of apsidal motion is
$ U = 360^\circ P_a/\dot{\omega} $. 
The nonlinear predictions, corresponding to the fitted parameters, are plotted as blue and orange curves for primary and secondary eclipses, respectively.

\begin{table*}
\centering
\caption{Apsidal motion elements of V1647~Sgr and V2283~Sgr, uncertainties in brackets. }
\label{t2}
\begin{tabular}{clllll}
\hline\hline\noalign{\smallskip}
Element      &  Unit          &   V1647~Sgr       &  V2283~Sgr     \\
\hline\noalign{\smallskip}
$T_0$        & BJD-24\,00000  &  41829.2523 (3)  &  13784.1605 (2)  \\
$P_s$        & days           &  3.28280010 (8)  &  3.47142150 (5)   \\
$P_a$        & days           &  3.28285098 (8)  &  3.47148337 (5)   \\
$e$          &  --            &  0.412 (1)       &  0.489 (1)     \\
$\dot{\omega}_{\rm obs}$ & deg $\:\rm{cycle^{-1}}$ 
                                 & 0.00557 (7)   &   0.00642 (8)  \\
$\dot{\omega}_{\rm obs}$ & deg $\:\rm{yr^{-1}}$  &  0.620 (7)   & 0.675 (8)    \\
$\omega_0$   & deg               &  203.5 (5)  &  315.8 (5)   \\
$U$          & years             &  580 (20)   &  533 (20)    \\ 
\hline\noalign{\smallskip}
$\Sigma w (O-C)^2$ & day$^2$    & $6.7\cdot10^{-4}$ & $3.0\cdot10^{-3}$  \\ 
\hline
\end{tabular}
\end{table*}


\section{Light curve and radial velocities analysis}
\label{lcrv}

Because the data obtained at the SAAO, FRAM, and Boyden observatories
 are less precise and non-continuous compared to the \T\ data, only the \T\ light curves were selected to fit the light curves of both systems. 
The other photometric data available to us were used mostly
for mid-eclipse time determination and the solution of apsidal motion. 
The completely covered light curves were routinely analyzed using the
{\sc Phoebe} computer code, developed by \cite{2005ApJ...628..426P}, see also \cite{2018maeb.book.....P}, which is a user-friendly implementation of the traditional Wilson-Devinney code \citep{1971ApJ...166..605W}. 
However, solving the light curves of eccentric binaries in {\sc Phoebe} is a rather challenging and time consuming task. To reduce the long computation time for eccentric orbits, the \T\ light curves were phased-binned to 300 points each.

The light elements, eccentricity, and apsidal motion rate have been adopted 
from the previous apsidal motion analysis (see Sect.~\ref{apsmotion}, Table~\ref{t2}). The minima times cover a longer time span, and these quantities are derived with higher precision. The other light curve parameters have been fitted: the luminosities, temperature of the secondary, inclination, and Kopal's modified potentials \citep{1959cbs..book.....K, 1979ApJ...234.1054W}. These potentials ($\Omega \sim R^{-1}$) contain contributions from the star, its nearby companion, the star's rotation about its axis, and the star's rotation in its orbit.    

\begin{figure}[t]
\centering
\includegraphics[width=0.5\textwidth]{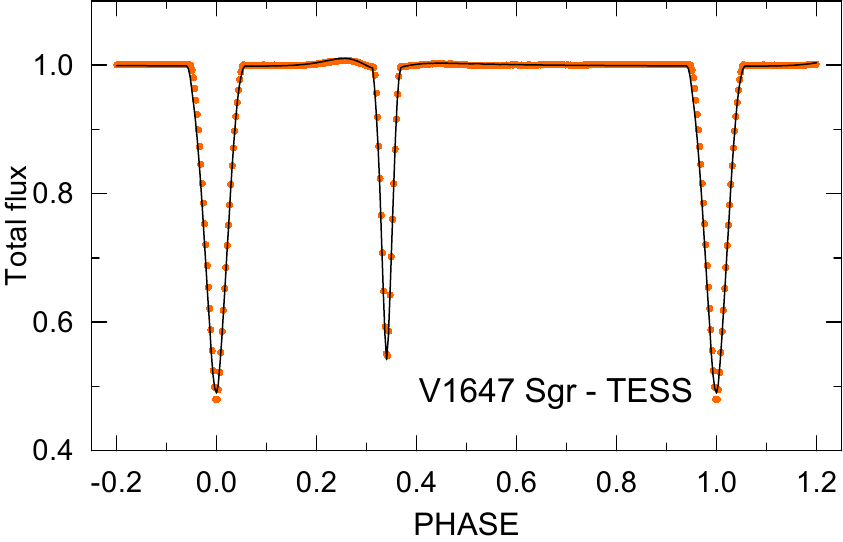}
\caption{\T\ light curve of V1647~Sgr obtained in Sector~66 (June 2023, orange dots, binning 300) and its {\sc Phoebe} solution (black curve). 
}
\label{1647lc}
\end{figure}

\begin{figure}[t]
\includegraphics[width=0.5\textwidth]{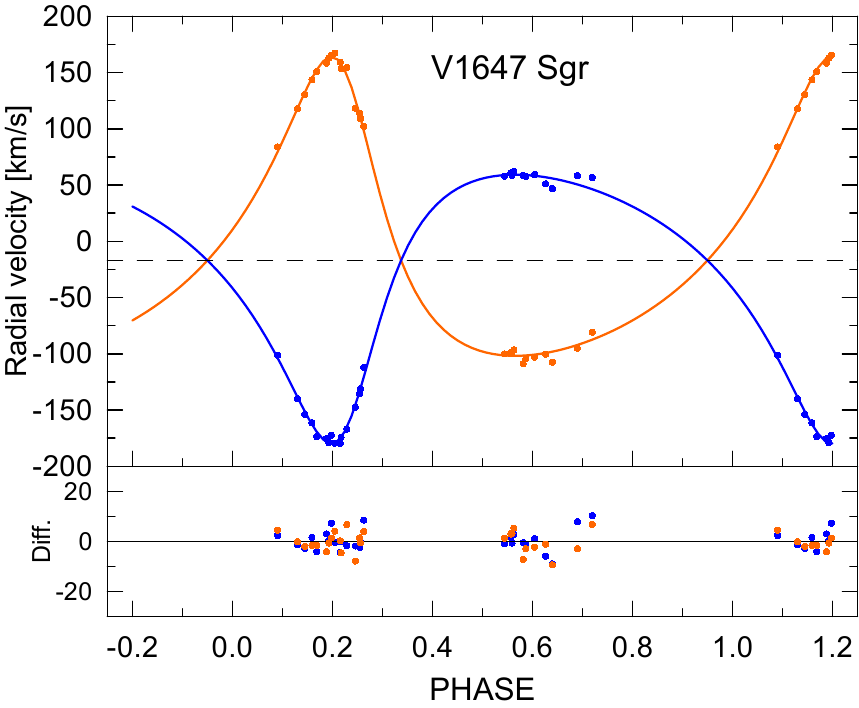}
\caption{Radial velocity curve for V1647~Sgr obtained by \cite{1985AA...145..206A} and its current solution in {\sc Phoebe}. Blue color denotes primary component, orange for secondary. The $\gamma$-velocity (--17 km/s) is plotted as a dotted line.
The residuals are given in the bottom panel.}
\label{1647rv}
\end{figure}

\begin{figure}[t]
\centering
\includegraphics[width=0.5\textwidth]{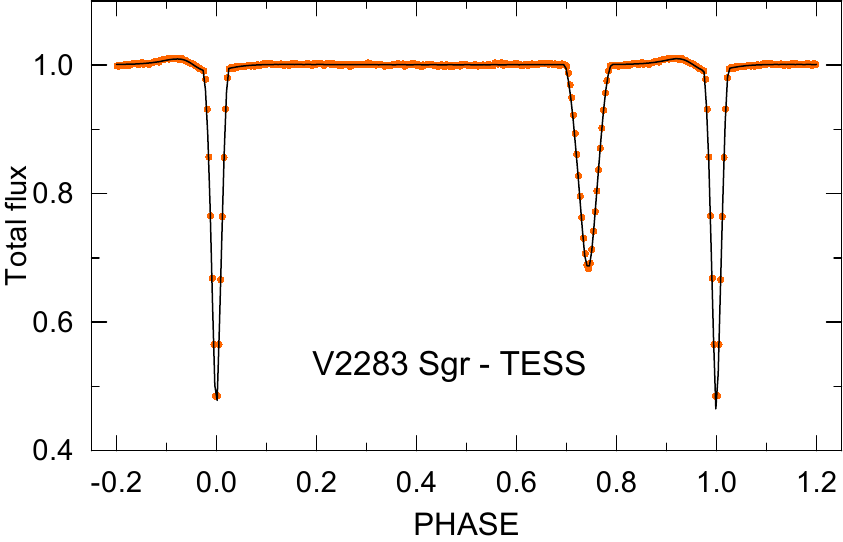}
\caption{\T\ light curve of V2283~Sgr with remarkable eccentricity obtained in Sector~39 (May/June 2021, orange dots, binning 300) and its {\sc Phoebe} solution (black curve). 
}
\label{2283lc}
\end{figure}

Because all components belong to early-type stars, we adopted bolometric albedos and gravity darkening coefficients as $A_1=A_2=1.0$ and $g_1=g_2=1.0$, which correspond to radiative envelopes.
We also assumed synchronous rotation for both components ($F_1 = F_2 = 1$).
Usually, in the case of elliptical orbits, the rotation tends to synchronize because of tidal interactions between the two components.
The limb-darkening coefficients were interpolated from van Hamme's
tables \citep{1993AJ....106.2096V}, using the logarithmic law.
We also tested the linear cosine limb darkening law and find little effect on our resulting parameters. For the solution of V1647~Sgr, the radial velocities of \cite{1985AA...145..206A} were used as an important input file.
The fine and coarse grid rasters for both components were set to 30.
The final solution was accepted when several subsequent iterations did not result in a decrease in the {\sc Phoebe} cost function.
The results of the photometric analysis are given in Table~\ref{lcfit}.
In this table, $T_1, T_2$ denotes the temperatures of the primary and secondary
components, $r_1, r_2$ the relative radii, $i$ the orbital inclination, and 
$\Omega_1$, $\Omega_2$ Kopal's modified potentials.
The corresponding light curves are plotted in Fig.~\ref{1647lc}, and \ref{2283lc} respectively.
The \T\ phased-binned light curve of both systems shows out-of-eclipse variations caused by the effect of light reflection from one component to the other close to the periastron and is well resolved using the {\sc Phoebe} code. 
The results of the simultaneous analysis of the light curve and the radial velocity curve, and the absolute dimensions for V1647~Sgr are listed in Table~\ref{PhysParam}. 
The radial velocity curve for V1647~Sgr obtained by \cite{1985AA...145..206A}  and its current solution are plotted in Fig.~\ref{1647rv}.  

For V2283~Sgr, different values of the temperature of the primary component can be found:
TESS Input Catalog \citep{2019AJ....158..138S} 8944 K,
{\it Gaia} BP/RP spectra \citep{2024AA...684A..29V} 9600~K,
Modern Mean Dwarf Stellar Color and Effective Temperature Sequence \citep{2013ApJS..208....9P}  9700 K,
 {\it Gaia} DR3 \citep{2022yCat.1355....0G}   9777~K.
The temperature $T_1 = 9700\pm 300$~K was adopted as a probable value based on these several sources. 
The spectral type and masses were then interpolated in the table of \cite{2013ApJS..208....9P}\footnote{A Modern Mean Dwarf Stellar Color and Effective Temperature Sequence \url{https://www.pas.rochester.edu/~emamajek/EEM_dwarf_UBVIJHK_colors_Teff.txt}}.
The temperature of the secondary and the dimensions of the components were then estimated using the {\sc Phoebe} solution. The semi-major axis $a$ in solar radii was computed according to the 3rd Kepler law.



\begin{table}
\caption{Parameters of the fit to the \T\ light curve in {\sc Phoebe.}}
\label{lcfit}
\centering
\begin{tabular}{lccc}
\hline\hline\noalign{\smallskip}
 Element           &   V1647~Sgr  & V2283~Sgr      \\
\hline\noalign{\smallskip}
 $T_1$ [K] (fixed) & 9\,600 (300)*  &  9\,700 (300) \dag  \\
 $T_2$ [K]         & 9\,000 (300)   &  7\,950 (300)   \\
 $r_1 = R_1/a$     &  0.1232 (45)   &  0.1201 (50)         \\
 $r_2 = R_2/a$     &  0.1150 (45)   &  0.1033 (50)         \\
 $e$ (fixed)       &  0.412         &  0.489         \\ 
 $i$ [deg]         &  89.9 (4)      &  89.5 (5)      \\
 $\Omega_1$        & 9.645   & 9.717    \\
 $\Omega_2$        & 8.927   & 8.986    \\
 \hline\noalign{\smallskip}
\end{tabular}

\medskip
Note: * value taken from \cite{2010AARv..18...67T}, \\
\dag~probable value according to literature.
\end{table}

\begin{figure}[t]
\centering
\includegraphics[width=0.45\textwidth]{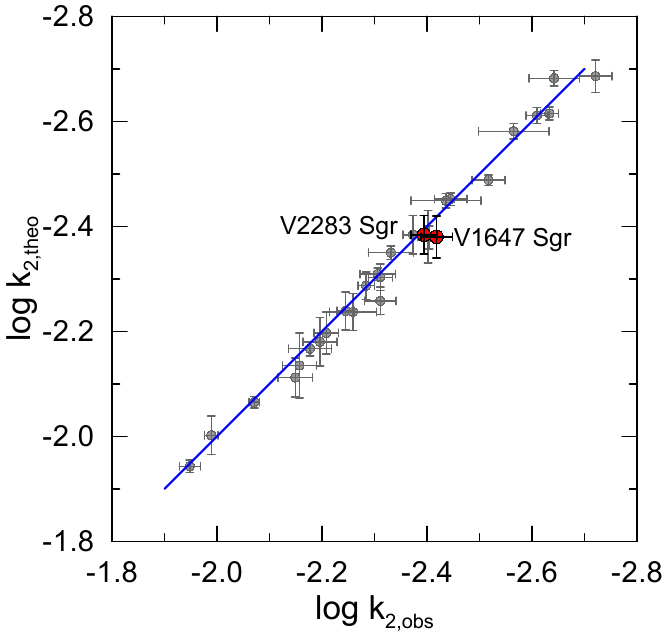}
\caption{Comparison of theoretical and observed internal structure constants 
 for V1647~Sgr and V2283~Sgr (red dots) with other known} eccentric systems 
collected in \cite[their Table~3] {2021A&A...654A..17C}. 
\label{k2comp}
\end{figure}

\begin{table*}
\centering
\caption{Astrophysical parameters and internal structure constants and their comparison.} 
\label{PhysParam}
\begin{tabular}{lccccccc}
\hline\hline\noalign{\smallskip}
Parameter &  Unit  &  V1647~Sgr &   V1647 Sgr        &  V2283~Sgr     \\
          &        & this paper & \cite{2010AARv..18...67T} & this paper  \\
\hline\noalign{\smallskip}
$M_1$   & \ms  &  2.184 (35)  &    2.184 (37)   & 2.178 (100)     \\
$M_2$   & \ms  &  1.957 (35)  &    1.967 (33)   & 1.547 (100)     \\
$R_1$   & \rs  &  1.839 (15)  &    1.832 (18)   & 1.796 (10)    \\
$R_2$   & \rs  &  1.716 (15)  &    1.667 (17)   & 1.544 (10)    \\
$\log g_1$ & cgs &  4.248       &  4.2517 (78)   & 4.267    \\
$\log g_2$ & cgs &  4.187       &  4.2879 (84)   & 4.250    \\
$q = M_2/M_1$ & --  & 0.899 (12) &  0.901 (9)    &   0.71     \\
$a$           & \rs &  14.92    &  14.33       &  15.0      \\
$\gamma$ & \ks   &  --16.97     & --16.8       &  --  \\  
\hline\noalign{\smallskip}
$\dot{\omega}_\mathrm{rel}$  & deg $\:\rm{cycle^{-1}}$ 
                             & 0.000\ 77 & -- & 0.000\ 75  \\
$\dot{\omega}_\mathrm{rel} / \dot{\omega}$ & \% &  13.8 & -- & 11.7  \\

log $k_\mathrm{2, obs}$  & -- & --2.394 (25) & --2.373 (19) & --2.418 (30) \\
log $k_\mathrm{2, theo}$ & -- & --2.384 \dag & --2.384 (37) & --2.38(4) \ddag \\
\hline
\end{tabular}

\medskip
Note: 
\dag~value taken from \cite{2021A&A...654A..17C},
\ddag~value interpolated in \cite{2023A&A...674A..67C} models for expected masses and age. 
\end{table*}

\section{Internal structure constant}
\label{ISC}

\begin{figure}[h]
\centering
\includegraphics[width=0.43\textwidth]{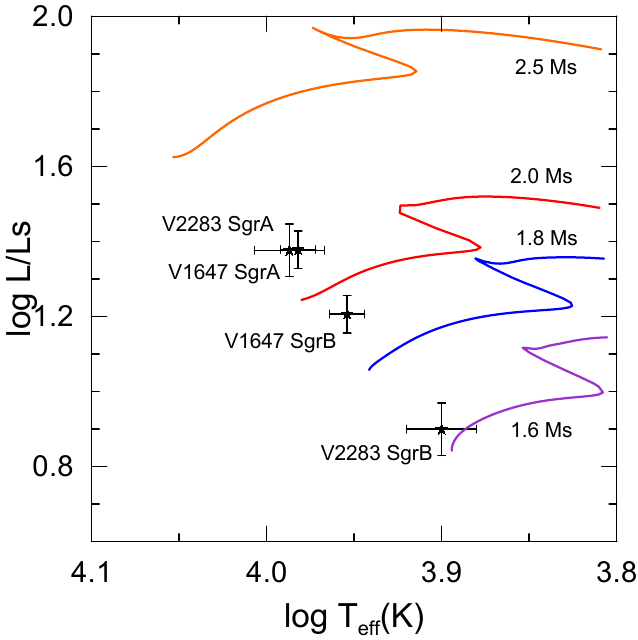}
\includegraphics[width=0.43\textwidth]{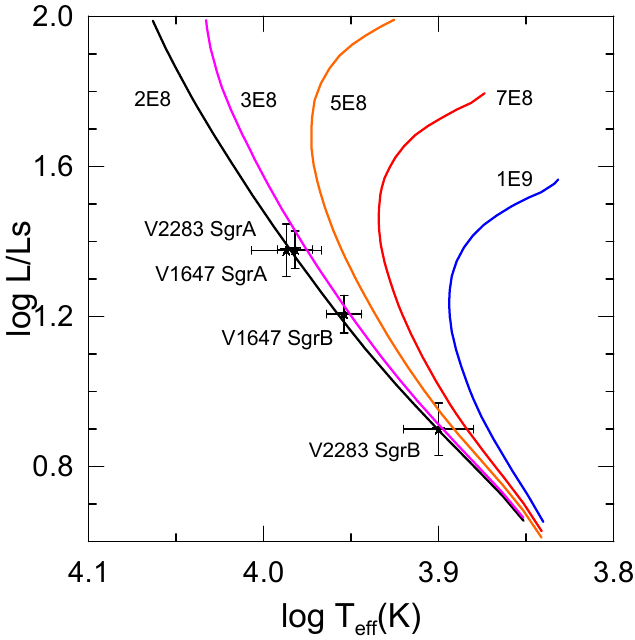}
\caption[]{HR~diagram for components of studied binaries. 
          Upper panel: Models of stellar evolution according to \cite{2019A&A...628A..29C} 
          for masses of 1.6, 1.8, 2.0, and 2.5 \ms\ are plotted. 
          Bottom panel: Isochrones calculated for ages from $2\cdot 10^8$ 
          to $10^9$ yr according to CMD~3.7 and PARSEC~2.0 models are given.
          The similarity of primary components of both binaries are clearly visible.}
\label{HRD}
\end{figure}

The great advantage of the detected apsidal motion in binaries is its direct connection with the internal structure constant (ISC), $k_2$, of both components. 
Generally, ISC is a measure of the density profile inside the star and is an important additional parameter of stellar evolution models. The mean value of the observed $\overline k_{2,\rm obs}$ is given by

\begin{equation}
\overline k_\mathrm{2, obs} = \frac{1}{c_{21} + c_{22}} \, \frac{P_a}{U}
         = \frac{1}{c_{21} + c_{22}} \, \frac{\dot{\omega}_{\rm cl}}{360} ,
\end{equation}

\noindent
where $c_{21}$ and $c_{22}$ are functions of the orbital eccentricity,
fractional radii,  masses of the components, and the ratio between the
rotational velocity of the stars and the Keplerian velocity \citep{1978ASSL...68.....K} 
and $\dot{\omega}_{\rm cl}$ represents the classical (or Newtonian) contribution to apsidal motion; see below.
The rotation of the stars was assumed to be synchronized with
the maximum angular orbital velocity achieved at periastron.

The observed ISC is very sensitive to radii of components ($k_\mathrm{2, obs} \sim R^5$),
and this test can only be performed for systems with accurate radii that provide
eclipsing binaries.
The theoretical internal structure constant $k_\mathrm{2, theo}$
is a combination of the ISC for both components

\begin{equation}
k_\mathrm{2, theo} = \frac{c_{21}k_{21} + c_{22}k_{22}}{c_{21} + c_{22}} , 
\end{equation}

\noindent
which can be compared with the observational value.

The observed apsidal motion rate has two independent components:
the classical (Newtonian) term, due to the non-spherical shape of both stars, 
and the relativistic term due to general relativity effects. 
Taking into account the value of the eccentricity and the masses of the components, 
a relativistic term, $\dot{\omega}_{\rm rel}$, must first be subtracted.
The original equation of \cite{1937AJMat..59..225L} could be rewritten
to be a more suitable function of known observable parameters, in degrees
per cycle \citep{1985ApJ...297..405G, 2023AJ....166..114D}:

\begin{equation}
\dot{\omega}_\mathrm{rel} = \; 5.447 \times 10^{-4} \: \frac{1}{1-e^2}
          \:\Biggl( \frac{M_1 + M_2}{P} \Biggr) ^{2/3},
\end{equation}

\noindent where $M_i$ denotes the individual masses of the
components in solar units, and $P$ is the orbital period in days.
It is clear that systems with higher stellar masses and shorter orbital
periods will have a larger relativistic apsidal motion.
The value of the classical contribution to the observed
apsidal motion rate, $\dot{\omega}_\mathrm{cl}$ is then the following:

\begin{equation}
\dot{\omega}_\mathrm{cl} = \dot{\omega}_\mathrm{obs}  - \dot{\omega}_\mathrm{rel}.
\end{equation}
\smallskip

\noindent
The astrophysical parameters of both systems are collected in Table~\ref{PhysParam},
as well as the values of $\dot{\omega}_\mathrm{rel}$ and the resulting mean internal
structure constants $\overline k_\mathrm{2,obs}$. Their errors were determined using
the relation derived in \cite{2005AA...437..545W}. 

\section{Discussion and conclusions}
\label{Concl}

This study provides actual information on the apsidal motion rates, absolute parameters, and observed ISC values for two close southern EEBs.
Our results for apsidal motion and light-curve analyzes of V1647~Sgr 
are similar to the parameters previously derived by \cite{2000AA...356..134W}
and \cite{2010AARv..18...67T}.  
This system consists of two similar stars of A1 spectral type. 
It was also confirmed that V1647~Sgr shows a slow but measurable
apsidal motion with a rather long period of about $580 \pm 20$ years.
The masses of both components $M_1$ = 2.184 \ms\ and $M_2$ = 1.957 \ms\ were slightly improved.
For the primary component of V2283~Sgr we found almost the same absolute parameters as for V1647~Sgr primary. The apsidal motion period in this system is somewhat shorter, $U = 533 \pm 20$ years. The secondary component is smaller and cooler.
The values obtained for the mean ISC, $\overline k_{\rm 2,obs}$ are compared
with their theoretical values $k_{\rm 2,theo}$ according to the current 
models along the main sequence computed by \cite{2023A&A...674A..67C}.
The agreement within the given errors can be seen in Fig.~\ref{k2comp}. 

The absolute dimensions of the components of both systems are given in Table~\ref{PhysParam}. 
The values for V1647~Sgr are sufficiently precise for comparison with current theoretical models, its masses and radii are known with an uncertainty of about 1-2\%. Their positions in the HR~diagram, which contains the evolutionary models for different masses according to \cite{2019A&A...628A..29C} are plotted in Fig.~\ref{HRD}.  
In the bottom part of this figure, the position of the components is compared with the set of isochrones calculated for ages from $2\cdot 10^8$ to $10^9$ years, according to the CMD~3.7 web interface and the PARSEC~2.0 models \citep{2012MNRAS.427..127B} available on the web pages of the Osservatorio Astronomico di Padova.~\footnote{\url{http://stev.oapd.inaf.it/cgi-bin/cmd\_3.7}.  }
The common age of all components is assumed to be about $2 \cdot 10^8$ years.

Since its discovery, a total of 22~positional measurements of V1647~Sgr have been published. These show only very negligible motion on the sky, indicating a very long orbital period of the double. Our calculations show that the orbit should be longer than 1000~years; hence, any significant movement of the double around a common barycenter is undetectable with our data (both in RVs of the components, as well as \oc\ of the eclipsing inner pair). Due to its separation in the sky (nowadays about 7.5$^{\prime\prime}$) and about 2~magnitudes lower brightness, this distant component is also not visible in the spectra of V1647~Sgr itself. Both of these components  share the same parallax and similar proper motion (according to the GAIA~DR3 catalogue), hence it is probably a classical hierarchical triple system. Quite remarkable is also the presence of another "C" component in the WDS (Washington Double Star Catalog). This component is distant about 90$^{\prime\prime}$, but its connection to V1647~Sgr is highly speculative. It shares a similar parallax; however, its proper motion is significantly different. Hence, we conclude that this component is probably not connected to V1647~Sgr itself.

Both systems, V1647~Sgr and V2283~Sgr are members of an important group of eclipsing binaries with precise absolute dimensions that are suitable for subsequent studies. 
Although the presence of a close third body is relatively common in eccentric binaries \citep{2017AcA....67..257W,2022NewA...9201708W}, it has not been demonstrated in these two systems, and no indications of the presence of a third component were observed in their \oc\ diagrams or light curves.
In the case of V2283~Sgr, we do not yet know the absolute parameters with great precision. It would also be desirable to obtain new high-dispersion and high-S/N spectroscopic observations to obtain the radial velocity curve and derive accurate masses for this system. Radial velocity amplitudes of about 200~km/s can be expected.

\medskip
\begin{acknowledgements}
Useful suggestions and recommendations from an anonymous referee helped us improve the clarity of the article and are greatly appreciated.
The research of MW and PZ was partially supported by the project
{\sc Cooperatio - Physics} of Charles University in Prague.
This publication was produced within the institutional support framework
for the development of the research organization at Masaryk University.
This investigation was supported by the allocation of SAAO observation time.
We thank the SAAO staff for their warm hospitality and assistance with the equipment.
The authors thank Jan Vraštil and Lukáš Pilarčík, former students at Astronomical Institute, Charles University Prague, for their important contribution to photometric observations with the DK154 telescope. 
This paper includes data collected by the \T\ mission. The funding for the \T\ mission is provided by the NASA Science Mission Directorate. 
Data presented in this paper were obtained from the Mikulski Archive for Space Telescopes (MAST).
The authors thank the Pierre Auger Collaboration for the use of its facilities. This work is supported by MEYS of Czech Republic under the projects MEYS LM2023032, LM2023047, and CZ.02.01.01/00/22\_008/0004632. The operation of the FRAM robotic telescope is supported by the grant from the Ministry of Education of the Czech Republic LM2018102. The data calibration and analysis related to the FRAM telescope is supported by the Ministry of Education of the Czech Republic MSMT-CR LTT18004, MSMT/EU funds CZ.02.1.01/0.0/0.0/16\_013/0001402, CZ.02.1.01/0.0/0.0/18\_046/0016010 and CZ.02.1.01/0.0/0.0/18\_046/0016007.   
The following Internet-based resources were used in the research for this paper:
the SIMBAD database \citep{2000A&AS..143....9W} 
and the VizieR catalog access tool \citep{2000A&AS..143...23O},
operated at CDS, Strasbourg Astronomical Observatory, France, 
NASA's Astrophysics Data System Bibliographic Services, 
the OMC Archive at LAEFF, preprocessed by ISDC, and
the VarAstro~\footnote{\url{https://var.astro.cz/en/}}, a portal for sharing photometric data by the Variable Star and Exoplanet Section, Czech Astronomical Society. 
This investigation is also part of an ongoing collaboration between professional astronomers and the Czech Astronomical Society, Variable Star and Exoplanet Section.
\end{acknowledgements}

\bibliographystyle{aa_link}
\bibliography{pceb} 

\begin{appendix}

\section{Observations and mid-eclipse times}

\renewcommand\thefigure{A.\arabic{figure}}

\begin{table*}
\caption {Selected local comparison and check stars at different observatories.}
\label{comp}
\begin{tabular}{clcl}
\hline\hline\noalign{\smallskip}
Variable & Comparison \& & $V$   & Used for photometry  \\
         & Check stars & [mag] &            \\
\hline\noalign{\smallskip}
V1647~Sgr & HD~162926*        &  6.053 &  PEP at SAAO in 2004-5 and 2019  \\ 
          & HD~164245*        &  6.296 &  PEP at SAAO in 2004-5 \\ 
          & UCAC4 263-150753 $\dag$ & 7.182  &  CCD at FRAM in 2020 \\
          & UCAC4 264-142401  & 8.340  &  CCD at FRAM in 2021 \\      
          & UCAC4 266-143028  & 10.34  &  CCD at Boyden in 2025\\
       
\hline\noalign{\smallskip}
V2283~Sgr & HD~321231          & 10.51  &   CCD at La Silla in 2017\\
          & UCAC4 266-150561   & 10.69  &   CCD at Boyden and FRAM in 2025  \\
          & UCAC4 266-149980   & 11.50  &   CCD at Boyden in 2025  \\
          & UCAC4 266-150489   & 10.73  &   CCD at FRAM in 2025 \\ 
\hline\noalign{\smallskip}
\end{tabular}

Notes: * used also by \cite{1977A&A....58..121C},
        $\dag$ suspected of variability
\end{table*}

\begin{table*}
\caption{New ground-based times of primary and secondary eclipses of V1647~Sgr and V2283~Sgr.}
\label{minground}
\begin{tabular}{lcclc}
\hline\hline\noalign{\smallskip}
BJD --     &    Primary/  & Epoch  & Error   &  Observatory    \\
24~00000   &    Secondary &        & [day]   &  Source   \\
\hline\noalign{\smallskip}
     \multicolumn{5}{c}{V1647~Sgr}  \\
     
53122.47841* & P & 3440.0 & 0.0001  & SAAO  \\
53444.1898   & P & 3538.0 & 0.001   & Pi of the Sky \\
53631.30885* & P & 3595.0 & 0.0001  & SAAO  \\
53632.30894* & S & 3595.5 & 0.0001  & SAAO  \\     
53809.58157  & S & 3649.5 & 0.0004  & \cite{2010IBVS.5931....1Z} \\
53903.77889  & P & 3678.0 & 0.0005  &  Pi of the Sky   \\
53983.5713	 & S & 3702.5 & 0.0010  & \cite{2008IBVS.5843....1O} \\
53983.5693   & S & 3702.5 & 0.001   &  VarAstro  \\
54353.51691  & P & 3815.0 & 0.002   &  Pi of the Sky   \\
54354.52841  & S & 3815.5 & 0.002   &  Pi of the Sky   \\
54652.25246  & P & 3906.0 & 0.0005  &  ASAS \\
54653.26823  & S & 3906.5 & 0.0005  &  ASAS \\
55385.338    & S & 4129.5 & 0.004   &  \cite{2010OEJV..130....1P} \\
57248.92319  & P & 4697.0 & 0.0002  &  \cite{2016JAVSO..44...26P}  \\
57501.69678  & P & 4774.0 & 0.002   &  \cite{2017OEJV..181....1P} \\
58681.29755* & S & 5133.5 & 0.001   &  SAAO  \\ 
58763.36898  & S & 5158.5 & 0.0005  &  OMC  \\
58975.65853  & P & 5223.0 & 0.001   &  FRAM   \\
59091.65077  & S & 5258.5 & 0.001   &  FRAM   \\ 
59415.54980  & P & 5357.0 & 0.0002  &  FRAM   \\
60840.26754* & P & 5791.0 & 0.0001  &  Boyden  \\
60841.39834* & S & 5791.5 & 0.0001  &  Boyden  \\
\hline\noalign{\smallskip}
   \multicolumn{5}{c}{V2283~Sgr} \\

52132.5608   & P &  11047.0 & 0.001   & \cite{2004IBVS.5542....1D} \\
53617.53614  & S &  11474.5 & 0.0001  & \cite{2018ApJS..235...41K} \\
53868.27954  & P &  11547.0 & 0.0001  & \cite{2018ApJS..235...41K} \\
57436.936    & P &  12575.0 & 0.005   & La Silla \\
60533.47890* & P & 13467.0 & 0.0001  & Boyden  \\
60821.61022* & P &  13550.0 & 0.0001  & Boyden  \\
60872.76733  & S &  13564.5 & 0.0001  & FRAM  \\
60873.68241  & P &  13565.0 & 0.0001  & FRAM  \\

\hline\noalign{\smallskip}
\end{tabular}
\smallskip

Note: * mean value of \textit{UBV, VR, VI} or {\it gri} measurements
\end{table*}

\begin{table}
\begin{center}
\caption{New \T\ times of primary and secondary eclipses of V1647~Sgr and V2283~Sgr derived in individual sectors.}
\label{mintess}
\begin{tabular}{ccc|cc}
\hline\hline\noalign{\smallskip}
TESS   & BJD --     & Epoch &  BJD --   & Epoch   \\
Sector & 24 00000   &       &  24 0000  &        \\
\hline\noalign{\smallskip}
 &  \multicolumn{2}{c}{V1647~Sgr}  &  \multicolumn{2}{c}{V2283~Sgr} \\
13 & 58660.51266 & 5127.0  &   58658.01711 &   12926.5 \\
 & 58661.60068  &  5127.5  &   58658.89177 &   12927.0 \\
 & 58663.79562  &  5128.0  &   58661.48825 &   12927.5 \\
 & 58664.88317  &  5128.5  &   58662.36318 &   12928.0 \\
 & 58676.92638  &  5132.0  &   58664.95986 &   12928.5 \\
 & 58678.01469  &  5132.5  &   58665.83471 &   12929.0 \\ 
 & 58680.20956  &  5133.0  &   58669.30585 &   12930.0 \\
 & 58681.29775  &  5133.5  &      --       &    --     \\
\hline\noalign{\smallskip}
39 & 59363.02430  &  5341.0  &  59362.71076  &  13129.5     \\
 & 59364.12574  &  5341.5  &    59363.59784  &  13130.0     \\ 
 & 59366.30682  &  5342.0  &    59367.06922  &  13131.0     \\ 
 & 59367.40867  &  5342.5  &    59370.54086  &  13132.0     \\
 & 59369.58978  &  5343.0  &    59373.12522  &  13132.5     \\
 & 59370.69145  &  5343.5  &    59377.48376  &  13134.0     \\
 & 59376.15546  &  5345.0  &    59380.06775  &  13134.5     \\ 
 & 59377.25718  &  5345.5  &    59380.95529  &  13135.0     \\ 
 & 59383.82263  &  5347.5  &    59383.53935  &  13135.5     \\
 & 59386.00367  &  5348.0  &    59384.42653  &  13136.0     \\
 & 59387.10551  &  5348.5  &    59387.01072  &  13136.5     \\
\hline\noalign{\smallskip}
66 & 60098.36333  &  5565.0  &  60116.00355  &  13346.5  \\  
 & 60099.47930  &  5565.5  &    60116.90441  &  13347.0  \\
 & 60106.04501  &  5567.5  &    60119.47481  &  13347.5  \\
 & 60108.21082  &  5568.0  &    60120.37585  &  13348.0  \\
 & 60109.32776  &  5568.5  &    60123.84738  &  13349.0  \\
 & 60111.49403  &  5569.0  &      --        &   --       \\
 & 60124.62507  &  5573.0  &      --        &   --       \\  
 & 60125.74183  &  5573.5  &      --        &   --       \\ 
\hline\noalign{\smallskip}
93 & 60830.41946 & 5788.0  &  --           &    --   \\  
 & 60831.54999 & 5788.5   &   60831.11044 & 13552.5  \\
 & 60833.70213 & 5789.0   &   60832.02502 & 13553.0  \\
 & 60834.83282 & 5789.5   &   60834.58184 & 13553.5  \\
 & 60836.98457 & 5790.0   &   60835.49652 & 13554.0  \\
 & 60838.11565 & 5790.5   &   60838.05323 & 13554.5  \\
 & 60840.26738 & 5791.0   &   60838.96796 & 13555.0  \\
 & 60843.55052 & 5792.0   &   60844.99611 & 13556.5  \\
 & 60844.68140 & 5792.5   &   60845.91089 & 13557.0  \\
 & 60846.83317 & 5793.0   &   60848.46745 & 13557.5  \\
 & 60847.96430 & 5793.5   &   60849.38236 & 13558.0  \\
 & 60851.24705 & 5794.5   &   60851.93893 & 13558.5  \\
 & 60853.39881 & 5795.0   &   60852.85381 & 13559.0  \\
 & 60854.52982 & 5795.5   &   --          &  --      \\
\hline
\end{tabular}
\end{center}
\end{table}

\end{appendix}
\end{document}